# On the Mechanism of Above Room Temperature Superconductivity and Superfluidity by Relativistic Quantum Mechanics

**Reginald B. Little**
**Department of Chemistry**
**Emory University**
**Atlanta, GA 30322**

## Abstract

A comprehensive theory of superconductivity (SC) and superfluidity (SF) is presented of new types III and IV at temperatures into millions of degrees involving phase transitions of fermions in heat reservoirs to form general relativistic triple quasi-particles of 3 interacting fermions to boson-fermion pairs. Types 0, I, and II SC/SF are deduced from such triples as: thermally dressed, relativistic fermionic vortices; spin coupled, dressed, fermionic vortical pairs (diamagnetic bosons); and spinrevorbitally coupled, dressed fermionic, vortical pairs (ferromagnetic bosons). All known SC, SF and trends in critical temperatures ($T_c$) are thereby explained. A theory of laminar to turbulent flow naturally arises out of this model. The recently observed SC/SF in nano-graphene systems is explained by this model. The above room temperature SC/SF is predicted and modeled by transformations of intense thermal boson populations of heat reservoirs to relativistic mass, weight, spin and magnetism for further reasoning over compression to electricity, weak phenomena and strong phenomena for connecting general relativism and quantum mechanics.

**Introduction**

Fermions in a heat reservoir may constitute many (very weak) energy momentum stress tensors (EMST) that can consolidate due to the strong magnetic field of the fermions and due to relative motions and random directions of thermal bosons to various stronger, denser vortical positions/momenta about the fermions for stronger, heavier superpositions (SP) of the fermions by general relativism of many bodies. Unlike a pure heat reservoir under substellar conditions, the mesoscopic bright and dark energies of the consolidating EMST of the fermions in the heat reservoirs do not completely repel to reform heat bosons of the reservoirs as the fermions can seed and contribute to the irreversible consolidations of the EMST by the Little Effect (1) to nucleate new weighty EMST, which in the limit of entraining synergistically more and more heat bosons contract under their own weights (due to magnetic frustration of rotation of the heat) into quantum, positional/momental fermionic vortices (FV) (Eqn 1) as the resulting tensor indices contract and lose continuity (due to time dilation, spatial contraction, spatial bending and eventual looping self-interactions) with cancellation of indicial terms (of mesoscopic bright and dark energies) except for a few indices as dictated by the nucleating (seeding) fermions in the reservoirs for the reduction of the EMST (16 partial differential eqns) of the compressed fermions and heat baths to Heisenberg matrices (Schrodinger and Dirac Eqns) of the positional vortices ($\Delta x$) and/or the momentum vortices ($\Delta p$) (orbitals, spin-orbitals and spinrevorbitals) as by the uncertainty principle (as the fermions lose their exact x and p in the quasifermions of the entrained thermal bosons) (wavelike to timelike) (nonlocal to local) with quantum entanglements and self-interactions of EMST of the heat generated vortices and the fermions for type 0 SC/SF as in $^3$He. The entrainment of heat by the fermions and transduction of heat to mass and then to sheer stress (magnetism) and pressure stress (electricity) by excess compression as reasoned here and required to justify quantum mysteries of superposition, entanglement and locality.

The many thermal bosons interact with the fermions more strongly and efficiently than photons due to general relativism of many relatively revolving 4 vectors of the thermal bosons of random directions with consequent heat influence on the dynamics of the fermions but photons are less influential on fermions as photons would require nonrelative revolutions of many coordinates. In the absence of strong magnetism, the fermions with many relative revolving thermal bosons leads to efficient thermal releases of heat from the fermions as by Kasha Rule. But in a strong magnetic field, the fermions orient, organize and synchronize relative to the many thermal bosons for their absorbance rather than release as by the Little Effect. Such magnetic organized absorbance of thermal bosons (momentum densities) of the heat bath initially forms gravitons and mass in the EMST for open spatial curvature, which with further Fourier thermal entrainment forms shear stress (magnetism) as the gravitons transform from open loops to closed loops of spatial curvature as the Fourier entrainment in the oriented thermal field shifts the open trajectories to closed quantum orbitals as described by Heisenberg matrix mechanics. So the magnetic fermions in the heat reservoir absorbs and transforms the thermal bosons by its magnetic field blocking relative revolutions of 4-vectors of thermal bosons by the Little Effect for many body Fourier heat entrainment in the magnetism and under greater weight with the time dilation, spatial contraction spatial bending and eventual looping self interactions for collapse of tensorial indices with many cancellations to form positional and momental matrices of Heisenberg. It is important to note that fermions can also entrain gravitons from other masses and they can entrain spacetime as they are displaced or move to excited triples and relativistic spinrevorbital (RSRO) continuum states which in the limit of low intensities such gravitons and spacetime are re-emitted from their continuum states to effect gravitational interactions and inertia of the fermions and even of bosons and triples and compositions thereof to be considered further within atoms and molecules of matter for quantum gravity.

As by this current theory, SC/SF are explained by motions exciting the triples into confined self-interacting magnetogravimetric states with self-interacting space loops (fermionic vortices) by collapsed EMST of stability without re-emissions and thereby without inertia. The EMST of entrained heat and consequent spacetime curvature for the vortical position about the fermion is given by:

1. $G_{\mu\nu,f}^{\text{position}} = (8 \pi G/c^2) * \sum_i^{\text{thermal}} (T_{\mu\nu}^{\text{fermion}})_i$

2. $m_{f,} = 8 \pi G/c^2 * (d\tau^2 / dx^\mu)^{\text{vortex}} \sum_i^{\text{thermal}} (T_{\mu\nu}^{\text{fermion}})_i^{\text{thermal}}$

3. $G_{\mu\nu,f}^{\text{position}} \cdot m_f \, dx_\mu / d\tau \; - \; m_f \, dx_\mu / d\tau \cdot G_{\mu\nu,f}^{\text{position}} = (ih/2\pi) \delta_{mn}$

From the various components (x,y,z,t) of EMST, the internal weight of such partially dressed FV in the heat bath is given by the curvature of spacetime by eqn (1) and the inertial mass of such dressed FV [as given by eqn (2)] is lowered relative to the gravitational weight by the factor $(d\tau^2 / dx^\mu)^{\text{vortex}}$ as the $\tau \to 0$, $dx_\mu \to \infty$ and vortical acceleration $(dx_\mu / d\tau^2) \to \infty$ for gravitational confinement and quantum gravity of the SC with loss of inertia for SF. The consequent space and momentum develop noncommutative properties due to spatial contraction and time dilation (eqn 3).

It is important to consider the evolution of the moving fermions in the heat bath and their entrainment of heat and transduction of the heat to mechanical, gravitational, electrical and magnetic energies with eventual collapse of the bath energy into superconducting and superfluid motion of the fermions. The collapse of the various EMTS of eqn (1) (and other eqns to follow) loses continuity in the limit of more and more spatial contraction-bending, time dilation and self-interactions such that only certain momenta and position (radii) for the Einstein Field Eqns [eqns (1,4,6,8,10,12,14, and 16)] are resonantly self-interacting and synchronized for forming discontinuum of vortices [eqn (3)] with intervening continuum (dissonant) vortical states. The consequent stable discontinuum momenta-positions (radii) cause noncommutation as by eqn (3). Surrounding fields (of heat, pressure, gravity, electricity and magnetism) of the bath can excite the FV discontinuum into continuum (commuting) states with re-emission of such perturbations to effect SF/SC as the perturbations into the continuum states produce huge transient gravity and magnetism between the shattered fermionic vortices for re-circulating the fragments back to SC/SF states. It is also important to note that above the $T_c$, the fermionic vortices are scattered by perturbations from phonons, dispersive forces, electric forces, and magnetic forces to manifest competing phases of square waves, stripe waves, charge density waves, pseudo gap states, magnetism and superconductivity. By this relativistic spinrevorbital mechanism and range of appropriate conditions, a theory of laminar to turbulence in flow can be explained as by considering at lower velocities the motion of fluid elements of fermions, bosons, and triples (as considered below) induce heat transformations and partial charging (by special theory of relativity) for interactions of fluid elements of a viscous type to determine laminar flow at low Reynolds numbers. As the velocity increases, this model explains a laminar to turbulent transition as by the general theory of relativity as the faster speed and higher fields locally bend space with the resulting local curvature disrupting fluid flow such that the layered flow of lamination is twisted locally with mixing locally and turbulent from the mixing at high Reynolds numbers.

At the faster speeds and higher fields the fermions acquire sufficient contraction of space and bending and time dilation so as to resonantly self-interact to form the discontinuum states of Eqn (3) and release waves and fields associated with non-allowed continuum dissonant states of the collapsed EMST. Eqn 3 expresses the resulting matrices of quantum mechanics of the position and momentum observed in the positional vortices. Gravitational quanta have to lose inertia and gain high velocities for quantum mechanics, but geometric alterations of space by phase transitions to self-interaction accommodate the dilemma of classic quanta lacking internal inertia and general relativism having gravitational-inertial equivalency. The consequent dressed fermions have more weight and acceleration against external electromagnetic drag forces and less internal inertia for its SC/SF. The Fourier entrainment of more external heat bosons due to greater internal weight with limitations by the decreasing magnetization of larger SP can form more positions ($G_i$) for variety of SP (G) for greater internal weight and lower internal inertia of the SP of vortices (G) as expressed from EMST in eqns 4 and 5 for SF/SF at higher temperatures in the heat reservoirs by heat transductions to SP by the fermion. The EMST of Fourier entraining heat for multiplex of positions (SP) is given by:

4. $G_{\mu\nu, f}^{SP} = (8 \pi G/c^2) * \sum_j^{vortices} (T_{\mu\nu}^{vortex})_j$

5. $m_f^{SP} = (8 \pi G/c^2) * (d\tau^2 / dx^{\mu})^{superpos} \sum_j^{vortices} (T_{\mu\nu}^{vortex})_j$

As the size of the SP increases, the magnetic field weakens and cannot prevent rotations of coordinates of the thermal bosons with less heat entrainment to limit size of SP. Thereby for type 0 SC/SF, the low energy thermal bosons of cryogenic temperatures are used to scatter the fermion among SC/SF states.

In addition to entangling heat in its magnetic field, under higher fermion densities the fermionic SP of vortices can entangle other FV in the heat baths to form bosons of various types. The fermions interact more strongly with each other than with the thermal bosons but the resulting bosons of type II due to their internal magnetism can entrain heat to form various SP of the bosons. Eqns 6 (7) and 8 (9) express the heavier internal weights and spacetime curvature (lower inertia) of bosons (relative to fermions) of type I (lacking internal relative motion) (with spin magnetic coupling and entanglement) and type II (having internal relative revolutions) (with orbital magnetic coupling and entanglement). The heavier weight of the dressed bosons relative to the dressed fermions allows more acceleration to counter electromagnetic drag forces of resistance for higher temperature SC/SF. Such bosons of type I can explain the SC and trends in $T_c$ for type I SC (Hg, Pb, Nb) with fermionic $e^- - e^-$ and nuclei – nuclei internal spin interactions. Figure 1. Bosons of type II can explain the SC and trends in $T_c$ for type II SC (NbN, $Nb_3Sn$, $Nb_3Ge$, and cuprates) with fermionic $e^-$ – nuclei, $e^-$ – $e^-$, nuclei – nuclei spin-orbital interactions. Figure 1. Such general relativistic FV also explains entanglement phenomena by spatial contraction/bending, time dilation and magnified interactions within the SP. Chemical phenomena of resonance, tautomerism and aromaticity are also captured by such general relativistic FV. The EMST of Fourier entraining heat for bosons of noninternal revolutions is given by:

6. $G_{\mu\nu,}^{boson, simult} = G_{\mu\nu, vortices}^{fermion SP} - G_{\mu\nu, vortices}^{fermion SP} = 0$

7. $m_{b, simul} = (F_{g, b, internal, simul}) (d\tau^2 / dx^{\mu}) = 0$

Thereby for type I SC/SF, $T_c$ is greater relative to type 0 due to internal spin magnetic coupling fermions to bosons. But the $T_c$ for type I is still low in order for heat bath to be unable to scatter the magnetically spin coupled fermions. The EMST of entraining heat for bosons of type II with internal revolutions is given by:

8. $G_{\mu\nu,}^{boson, nonsimult} = (8 \pi G/c^2) * \sum_k^{vortices, thermal} (T_{\mu\nu}^{vortex})_k^{boson}$

9. $m_{b, nonsimul} = (8 \pi G/c^2) * (d\tau^2 / dx^{\mu})^{boson} \sum_k^{vortices} (T_{\mu\nu}^{vortex})_k^{boson}$

Thereby for type II SC/SF, the $T_c$ is greater relative to type I due to ferromagnetic coupling of fermions of bosons due to internal motions of bosons. But the $T_c$ for type II is still low enough to prevent the heat bath from scattering orbitally magnetic coupled fermions.

Under even higher fermionic densities, the fermionic SP of vortices can Fourier entangle bosons to form triples of various types. Eqns 10 (11) and 12 (13) express the heavier internal weights (lower inertia) of triples (relative to bosons and fermions) of type III (lacking internal bosonic relative motion) and even heavier weights and lower inertia of triples of type IV (having internal relative bosonic motion). The greater weights and lower inertia of the dressed triples relative to dressed bosons allow more acceleration against electromagnetic drag for SC/SF at higher temperatures. The fermions interact with bosons more strongly (to form triples) than fermion-fermion interactions (to form bosons) for greater stability of triples at higher temperatures and energies relative to bosons. Such triples of type III can explain the SC and SF and trends in $T_c$ in various substances by including Fourier fermionic/bosonic entanglements for triples between both nuclei and electronic lattices ($RbCsC_{60}$, $NgB_2$, cuprates, and FeAs). See Figure 1. It is important to consider a weight factor (W) which is the potential energy (PE) ratio ($e^-$--$e^-$/$e^-$-nuclear) for coupling nuclear and electronic fermions and bosons for affecting relative weights and inertia across the Born-Oppenheimer limit. Larger W factors cause higher $T_c$ of the various SC in Figure 1 for explaining the trends. In the next section, recent observations of above room temperature SC and SF in graphene (GR) and nanosolution (NS) are related to type III triples. The type IV SC/SF may explain motion in extreme energetic environments like nuclei and stars. The EMST of Fourier entraining heat for triples of type III of noninternal bosonic revolutions is given by:

10. $G_{\mu\nu, \, t, \, simul} \sim (8 \pi G/c^2) * \sum_l^{rev} (T_{\mu\nu}^{vortices})_l^{triple}$

11. $m_{t, \, simult} = (8 \pi G/c^2) \, (d\tau^2 / dx^\mu)^{triple} \, \{\sum_l^{rev} (T_{\mu\nu}^{vortices})_l^{triple}\}$

The EMST of Fourier entraining heat for triples with internal bosonic revolutions is given by

12. $G_{\mu\nu \, t, \, nonsimul} = (8 \pi G/c^2) * \sum_l^{rev} (T_{\mu\nu}^{vortices})_l^{triple}$

13. $m_{t, \, nonsimul} = (8 \pi G/c^2) (d\tau^2 / dx^\mu)^{triple} * \sum_l^{rev} (T_{\mu\nu}^{vortices})_l^{triple}$

Thereby for type III and IV SC/SF, the thermal bosons are converted to magnons and quasifermions for higher $T_c$ of type III and IV relative to type II SC/SF as more thermal energy of the surrounding lattice (heat bath) is Fourier converted to magnetogravimetric (MG) binding of the SC/SF phase.

Due to the stronger interactions of dressed triples relative to dressed bosons and dressed fermions, the dressed triples form SC/SF phases at higher temperatures. The Fourier entanglement of fat weighty, multiplex, vortical triples to form SC/SF phases by the superpositions contributes to the augmenting of weights (and lowering inertia) by summing, multiplying and even exponentiating the many stress tensors due to $N_o$ (Avogadro order) relative speeds and directions of thermal bosons in the heat reservoir by more Fourier entrainment of ($N_o$) thermal bosons and triples with consequent internal contractions of the tensors to Heisenberg matrices. The EMST of Fourier entraining heat for triple-triple interactions to form phases of triples with noninternal bosonic revolutions is given by:

14. $G_{\mu\nu, ph, t \, simult} = 8 \pi G/c^2 * \sum_n^{rev} (T_{\mu\nu}^{vortices})_n^{phase}$

15. $m_{t, ph, \, simult} = 8 \pi G/c^2 * (d\tau^2 / dx^\mu)^{phase} * \sum_n^{rev} (T_{\mu\nu}^{vortices})_n^{phase}$

The EMST of Fourier entraining heat for triple - triple interactions to form phases of triples with internal bosonic revolutions is given by:

16. $G_{\mu\nu,ph,t,}{}^{nonsimult} = 8\pi G/c^2 * \sum_n{}^{rev}(T_{\mu\nu}{}^{vortices})_n{}^{phase}$

17. $m_{t, nonsimult} = 8\pi G/c^2 * (d\tau^2/dx^\mu)^{phase} * \sum_n{}^{rev}(T_{\mu\nu}{}^{vortices})_n{}^{phase}$

The greater W factors tend to increase $e^-$-$e^-$ kinetic energies and weights to enlarge $T_c$. Such effects of W factor explain $T_c$ trends for electronic lattices of various elements $T_c(N) < T_c(C)$; $T_c(N) < T_c(As)$; $T_c(Si) < T_c(Ge)$; $T_c(N) < T_c(O) < T_c(C), T_c(B)$. See Figure 1. In complex ionic structures, W factors increase by transfer of $e^-$ to polyanionic lattices and diminished $e^-$ - nuclear PE on $e^-$ - lattices; the consequent cations have diminished $e^-$ - nuclear PE for their ease of SF revolutions and Coulombic decoupling but magnetic coupling of cations to electronic lattices. $e^-$ lattices of different subshell hybridizations also raise W factors for higher $T_c$ due to $e^-$ of different velocities for greater densities and weights. Huge weight factors in GR-NS are reasoned for their SC/SF at above room temperature as given in results with discussion.

**Results and Discussion**

On the basis of the model, the MG, many-body Fourier interactions can form heavy fermions, heavy bosons, heavy triples and weighty phases of triples of layered domains as illustrated in Fig 2 and modeled in eqns 1-17. The layered symmetry and structures of the heat bath allow Fourier fermion – thermal boson, fermion-fermion, fermion – boson interactions of many EMST without rotations of coordinates for entrainment, consolidation, and collapse of tensors to matrices for SC/SF of ions and charges in and about the heat bath. Such layered SC/SF systems are manifested by SF aqueous nanosolutions (NS) of various salts transporting through GR oxide nanomembranes (2,3) and SC in nano GR - NS (4-6). The SC/SF in these GR systems are of type III and modeled by tensorial relationships in eqns 1-17 whereby the fermionic and bosonic nuclei of the cations { $p^+$, $Na^+$, $Mg^{2+}$, $Ba^{2+}$, $Ca^{2+}$, and $K^+$} and fermionic chloride anions (and $2p^+ + OH^-$) Fourier entrain thermal energies to form complex vortices with coupling to electronic vortices in the GR nanowalls (for type III SC/SF) at different temperatures of 20°C in Fig 3 and 10, 20, 30, and 40 °C in Fig 4 with greater EMST of heavier ions and bosonic ions { $Ba^{2+}$, $Ca^{2+}$, and $K^+$} (via greater energetic RSRO vortices) Fourier entraining (dressing) greater quanta of thermal energies at higher temperatures and with smaller EMST of smaller mass ions { $p^+$, $Na^+$, and $Mg^{2+}$} (and fermions via lesser energetic RSRO vortices) Fourier entraining (dressing) thermal energies in excess of their phase compressibilities (for the instability of SF of NaCl(aq) at temp > 20 °C) for transducing entrained thermal energies (back and forth motions) to bright and dark energies to bright and dark particles with consequent relativistic weights (eqns 1, 4, 6, 8, 10, 12, 14, and 16), gravitational energy, vortical quasi-fermions and magnetism within/about the layered phases for MG binding and quantum gravity in the layered phases over macroscopic spacetime and above room temperature for consequent SF of the layered, aqueous NS and multidomain aqueous macrosolutions by less of viscous, vibrational, Coulombic interactions of the aqueous NS layers with the surrounding type III SC - GR interface by gain of nonviscous, revolutionary, MG interactions between the type III SF aqueous NS and type III SC - GR nanowalls for further rationalizing greater compression, transforming magnetism to electricity (and ultracompression transducing electricity to weak phenomena and strong phenomena as by type IV triples in nuclear, earth's core and stellar materials).

In Fig 4, the heavier salt solutions require higher temperatures of the heat bath to organize SF of their solutions and couple to the SC in the nano GR walls. By magnetic diminished rotation of coordinates Fourier entraining thermal energy, cooling, gravitating, magnetizing, and compressing within/between the aqueous NS and GR nanopores, the EMST of the fermions and heat bosons in the GR – aqueous NS determine tensorial relations of eqns 1-17, which reduce into Heisenberg matrices of quantum mechanics (for the small flakes in Fig 3) by spatial contraction and time dilation and self-interactions due to the tensorial indices collapsing and contracting due to $\Delta x$ and $\Delta p$ changing as space contracts and time dilates with indices becoming restricted in values for quantization (discontinuum) rather than unrestricted (continuum) indices as curvature ($G_{\mu v, ph, t\ simult}$) no longer equals EMST ($T_{\mu v}^{vortices}$) and their products noncommute as unrestricted indices manifesting relativistic unobservable motions for a new way of connecting general relativism to quantum mechanics. Motion creates space but the space does not destroy the motion but perpetuates of order Planck's constant. The big flakes in Fig 3 have bigger graphene pores with less collapse of tensors to Heisenberg matrices and less SC/SF of the solutions and GR walls. By such connection many mysteries of quantum mechanics (SC, SF, superposition, entanglement and gravity [itself]) are explained as manifested in the GO membrane.

A complete solution of Einstein Field Eqns [7] within the aqueous nano solutions and the GR nano-pores would give complete description, but the expressions of eqns 1-17 capture the effects as in general the gravitational weights and inertial masses of layers within the aqueous NS and GR nanopores in Fig 2-4 increase and decrease (respectively) for greater MG binding and lesser inertial mass for complete prediction and explanation of the type III SF and SC in the GO nanomembranes. The theory here is magnetic and many body as opposed to the two body reasoning in ref 3. By this new model, the weights (inertia masses) increase (decrease) within and between layers due to general relativistic faster increase of $dx^\mu$ and relative to $d\tau$ and decreasing ($d\tau^2/dx^\mu$) for complete curvature to loops; whereby the smaller ($d\tau^2/dx^\mu$) diminishes inertia mass ( weight * $d\tau^2/dx^\mu$) for reversing the inertial trend relative to the weight trend breaking gravitational inertial equivalence over larger space-time with ionic enhancement and mixed subshell enhancement (hybrid orbitals). Moreover it is important to note that for the systems of Figures 2, 3, and 4, the MG exchange and correlation determine more complex relations for exponential rather than multiplicative functionalities of increase weight verses $dx^\mu$ and $d\tau$. Thereby for exponential functionalities even lower velocities (relative to c), smaller $dx^\mu$ (less spatial contractions and time dilations) and larger $d\tau$ (less distortion), all for magnified gravity and diminished inertia for SC and SF above room temperature.

**Acknowledgement**
I thank GOD.

**Figure 1 – Time Line of Superconductor Discoveries (credit: Wikipedia)**

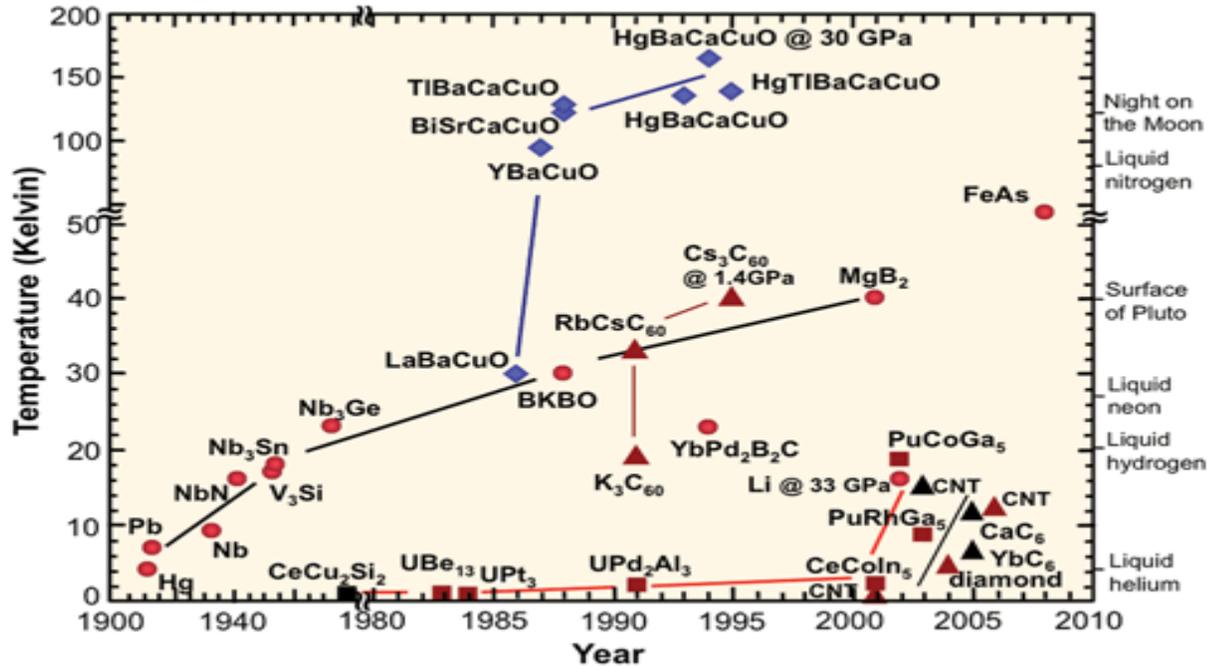

**Figure 2 – The Formation of Transient Relativistic Coupled Triples by Fermions and Bosons for Layers and Superfluidity and Superconductivity**

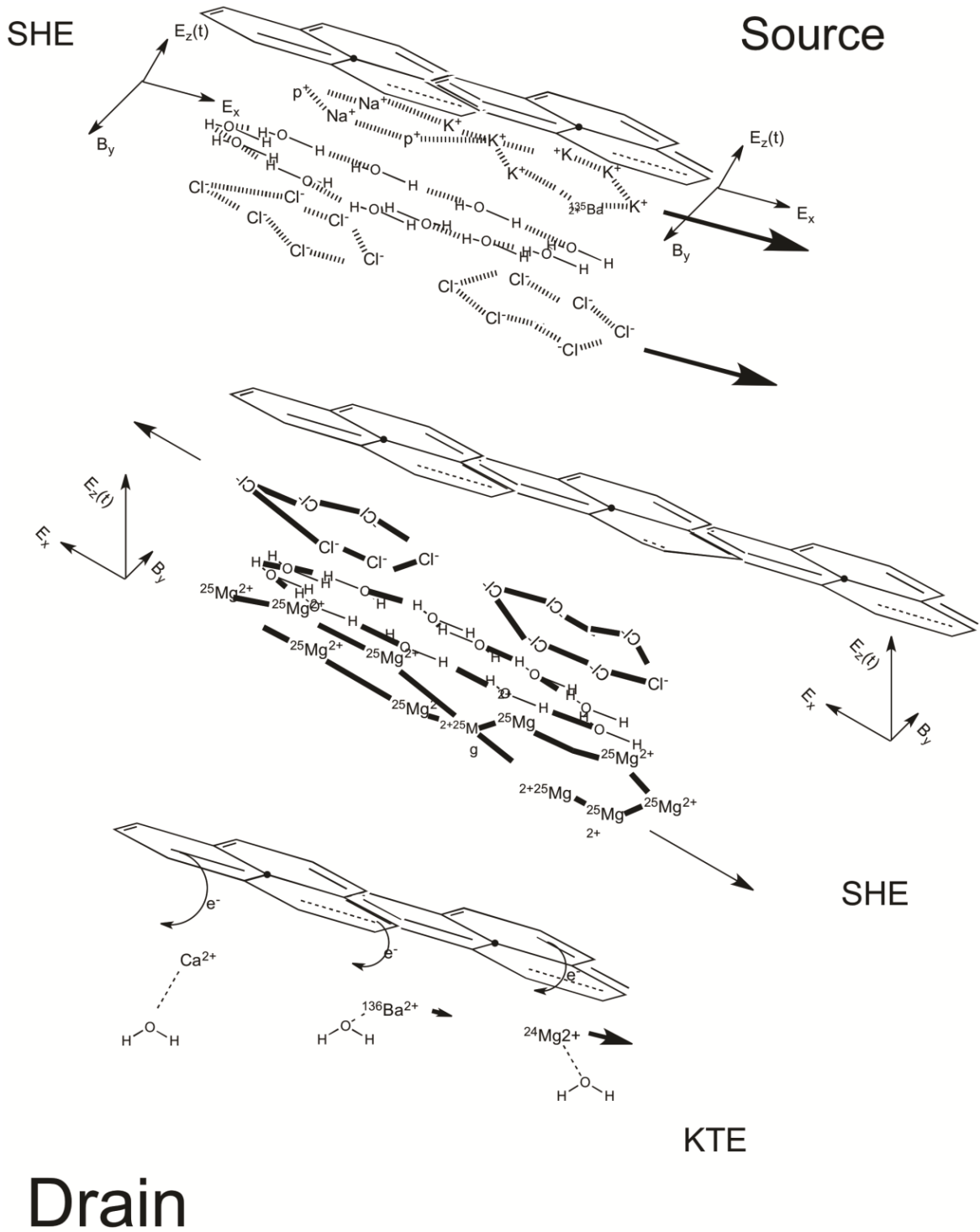

**Figure 3 - The Transport Data for Less Massive Alkali and Alkaline Earth Chloride Aqueous Nanosolutions Through GR/GO Nano-Membrane (Data regraphed from reference 3)**

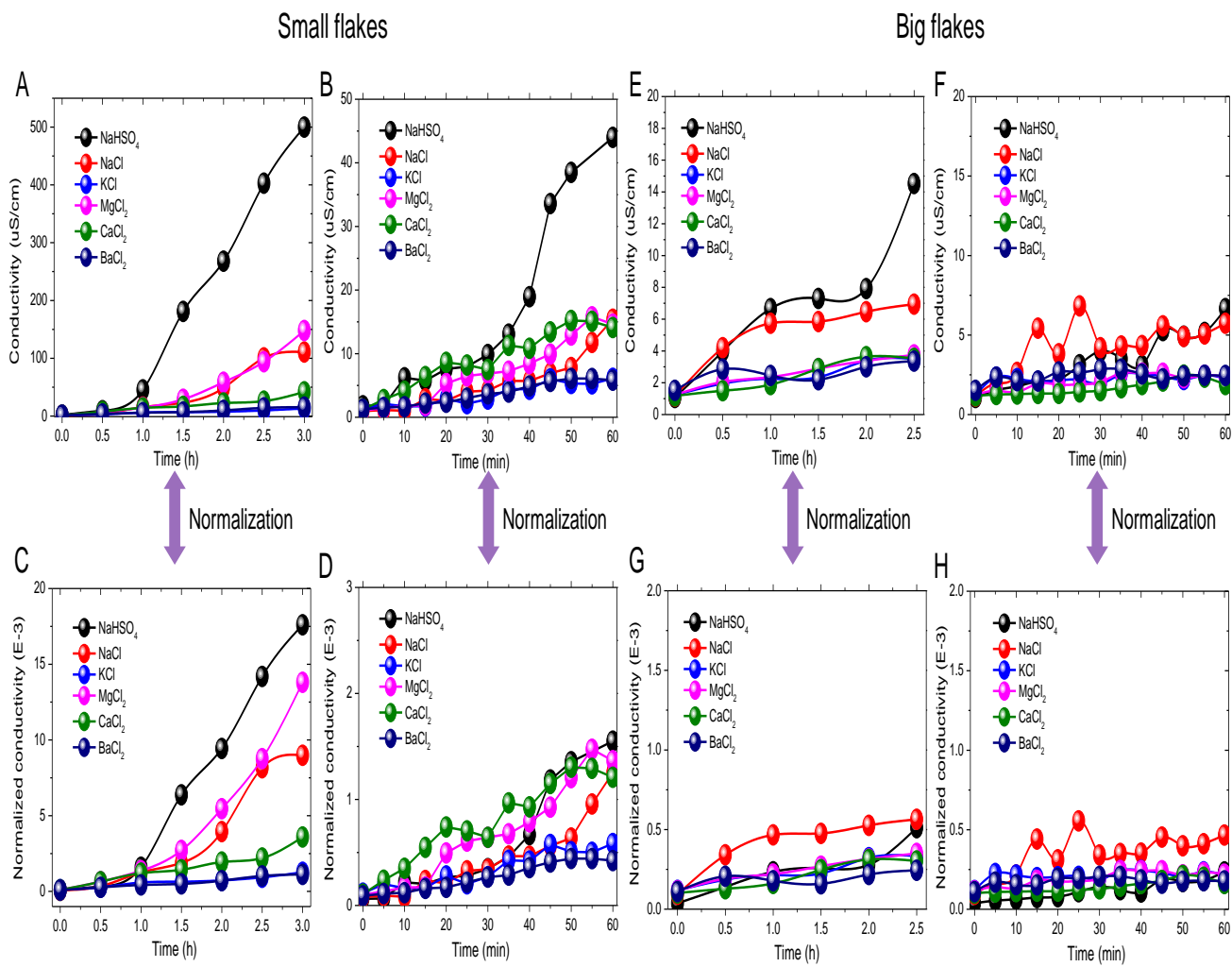

**Figure 4 – The Transport Data for Heavier Alkali and Alkaline Earth Chloride Aqueous Nanosolutions Through GR/GO Nano-Membrane. (Data regraphed from reference 3).**

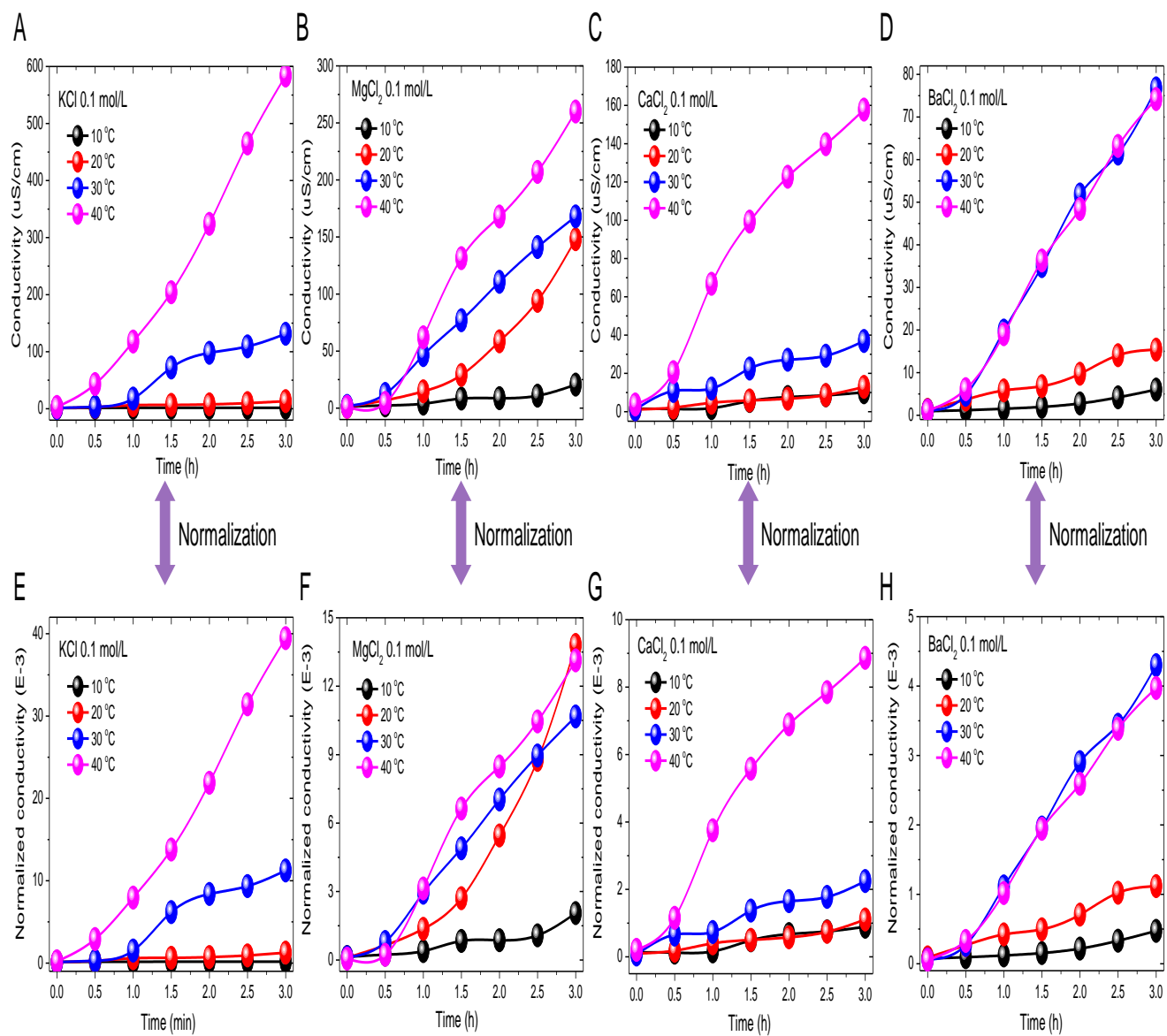